\documentclass[10pt,conference]{IEEEtran}\IEEEoverridecommandlockouts

\usepackage{CJK}
\usepackage{amsfonts}
\usepackage{slashbox}
\usepackage{amsmath}
\usepackage{multirow}
\usepackage{graphicx}
\usepackage{amsfonts}
\usepackage{amsmath,epsfig}
\usepackage{mathbbold}
\usepackage{stfloats}
\usepackage{amssymb}
\usepackage{url}
\usepackage{bm}
\usepackage{cite}
\usepackage{amsmath}
\usepackage{amssymb}
\usepackage{latexsym}
\usepackage{multirow}
\usepackage{epsfig}
\usepackage{graphics}
\usepackage{mathrsfs}
\usepackage{algorithmic}
\usepackage{algorithm}
\usepackage{subfigure}
\usepackage{slashbox}
\usepackage[all]{xy}

\usepackage{pifont}
\usepackage{bbding}
\usepackage{stmaryrd}
\usepackage{amssymb}
\usepackage{amsfonts}
\usepackage{epic}
\usepackage{stfloats}
\usepackage{latexsym}
\usepackage{epstopdf}
\usepackage{epic}
\usepackage{bm}
\usepackage{xcolor}
\usepackage{mathrsfs}
\usepackage{pifont}
\usepackage{bbding}
\usepackage{amsmath,epsfig}
\usepackage{mathbbold}
\usepackage{stmaryrd}
\usepackage{amssymb}
\usepackage{amsfonts}
\usepackage{epic}
\usepackage{graphicx}
\usepackage{subfigure}

\usepackage{enumerate}

\usepackage{stfloats}
\usepackage{latexsym}
\usepackage{epstopdf}
\usepackage{epic}
\usepackage{multirow}
\usepackage{stfloats}
\usepackage{bm}

\usepackage{booktabs}
\usepackage{color}
\usepackage{cases}

\newcommand{\aaa}{\bm{a}}

\usepackage{geometry}
\makeatother

\begin{document}
\title{{Intelligent Reflecting  Surface Enhanced Wireless Network: Joint Active and Passive Beamforming Design}}

\author{\IEEEauthorblockN{Qingqing Wu and Rui Zhang}
\IEEEauthorblockA{ Department of Electrical and Computer Engineering, National University of Singapore}
Emails: \{elewuqq, elezhang\}@nus.edu.sg}
\maketitle
\begin{abstract}
Intelligent reflecting surface (IRS) is envisioned to have abundant applications in future wireless networks by smartly reconfiguring the signal propagation for performance enhancement. 
Specifically, an IRS consists of a large number of low-cost passive elements each reflecting the incident signal with a certain phase shift to collaboratively achieve beamforming and suppress interference at one or more designated receivers.
In this paper, we study an IRS-enhanced point-to-point multiple-input single-output (MISO)  wireless system where one IRS is deployed to assist in the communication from a multi-antenna access point (AP) to a single-antenna user. As a result, the user simultaneously receives the signal sent directly from the AP  as well as that reflected by the IRS.
We aim to maximize the total received signal power at the user by jointly optimizing the (active) transmit beamforming at the AP and (passive) reflect beamforming by the phase shifters at the IRS. 
We first propose a centralized algorithm based on the technique of semidefinite relaxation (SDR) by assuming the global channel state information (CSI) available at the IRS. Since the centralized implementation requires excessive channel estimation and signal exchange overheads, we further propose a  low-complexity distributed algorithm  where the AP and  IRS independently adjust the transmit beamforming and the phase shifts in an alternating manner until the convergence is reached.
Simulation results show that significant performance gains can be achieved by the proposed algorithms as compared to benchmark schemes.
Moreover, it is verified that the IRS is able to drastically enhance the link quality and/or coverage over the conventional setup without the IRS.

\end{abstract}


\section{Introduction}  
Although there has been a quantum leap in spectrum efficiency of wireless networks in the last few decades thanks to various technological advances such as ultra-dense network (UDN), massive multiple-input multiple-output (M-MIMO), and  millimeter wave (mmWave) communications, the network energy consumption and hardware cost are still critical issues faced in practical implementation  \cite{zhang2016fundamental}. For example, UDNs almost linearly scale the circuit and cooling energy consumption with the number of newly deployed  base stations (BSs) \cite{wu2016energy}, while costly radio frequency (RF) chains  and complex signal processing techniques are needed for efficient communication at mmWave frequencies.  On the other hand, adding an excessively large number of  active components such as small-cell BSs/relays in wireless networks also causes a more severe interference issue \cite{wang2017joint}. Therefore,  research on finding both spectral and energy efficient techniques with low hardware cost is still imperative for realizing sustainable and green fifth-generation (5G)  wireless networks and beyond \cite{wu2016overview}.

In this paper, intelligent reflecting surface (IRS) is proposed as a promising green and cost-effective solution to achieve the above challenging goals.  
Specifically, IRS is a planar array consisting of  a large number of passive elements (e.g., low-cost printed dipoles), where each element is able to induce a certain phase shift  (by a smart controller) independently on the incident electromagnetic wave. As the key component of conventional reflectarrays, passive reflecting surface has found a variety of applications in radar and satellite communications, which,  however, is rarely used in terrestrial wireless communication. This is because traditional reflecting surfaces only have fixed phase shifters once  fabricated, which are difficult to meet the dynamics of wireless networks with time-varying channels. However, recent advances in RF micro electromechanical systems (MEMS) and metamaterial  (e.g., metasurface)  have made the reconfigurability of reflecting surfaces possible, even via controlling the phase shifters  in real time \cite{cui2014coding}. By smartly adjusting the phase shifts of all elements at an IRS, the reflected signals  can add coherently at the desired receiver to improve the received signal power or destructively at the non-intended receiver to avoid interference and enhance security/privacy. 

It is worth noting that  the proposed IRS differs significantly from other existing related technologies such as amplify-and-forward (AF) relay, backscatter communication, and active intelligent surface based M-MIMO \cite{hu2017beyond}.
First,  compared to AF relay that assists in source-destination transmission by actively generating new signals, IRS does not use a transmitter module but only reflects the ambient RF signals as a passive array, which thus incurs no additional power consumption.
Second, different from the  traditional backscatter communication of the radio frequency identification (RFID)  tag that communicates with the receiver by reflecting the incident wave sent from the reader, IRS is utilized  to enhance the existing communication link performance instead of delivering any information of its own. As such,  the direct-path signal (from reader to receiver) in backscatter communications is the undesired interference and hence needs to be canceled/suppressed at the receiver.  However, in IRS-enhanced communications, both the direct-path and reflect-path signals carry the same useful information and thus should be  coherently added at the receiver to maximize the total received power. Third, IRS is also different from the active intelligent surface based M-MIMO due to their different array architectures (passive versus active) and operating mechanisms (reflect versus transmit). Furthermore, IRSs possess other advantages, such as low profile, lightweight, and conformal geometry, which enable them to be easily attached/removed on/from the wall or ceiling, thus providing high flexibility and superior compatibility for practical implementation \cite{subrt2012intelligent}.
For example, by installing  IRSs on the walls which are in line-of-sight (LoS) of an access point (AP)/BS,  its signal strength and coverage are anticipated to be significantly improved.
All the above merits render IRSs an appealing solution for performance enhancement in future generation wireless networks, especially for indoor applications with a high density of users in e.g. stadiums, shopping malls, exhibition centers, and airports. However, the research on IRS design and performance optimization is in its infancy and there has been very limited work in this new area to the authors' best knowledge.
\begin{figure}[!t]
\centering
\includegraphics[width=0.450\textwidth]{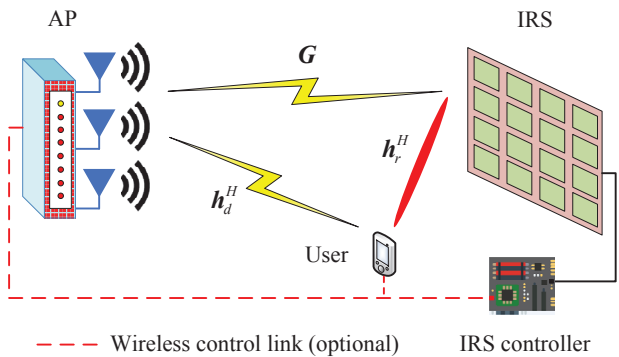}
\caption{An IRS-enhanced  wireless system. } \label{system:model}\vspace{-0.6cm}
\end{figure}

In this paper, we consider an IRS-enhanced wireless system as shown in Fig. \ref{system:model},  where a multi-antenna AP serves a single-antenna user with the help of an IRS (e.g., on the wall). Such a system can be employed to facilitate wireless information and/or power transfer in various Internet-of-things (IoT) applications \cite{wu2016overview,zhangrui13_mimo,qing15_wpcn_twc,wu2018spectral,xu2014multiuser,zhang2018wireless,abeywickrama2018wireless}.  Since the user receives the superposed signals from both the AP-user (direct) link and IRS-user link,
 we jointly optimize the (active) transmit beamforming at the AP and (passive) reflect beamforming by the  phase shifters at the IRS to maximize the total signal power received at  the user. Intuitively,  if the channel of the AP-user link is much stronger than that of the AP-IRS link, it is preferable for the AP to beam toward the user directly, while in the opposite case, especially when the AP-user link is severally blocked by e.g., corridors, as often encountered  in indoor  applications, the AP would adjust its beamforming direction toward the IRS to leverage its reflected signal to serve the user.  In this case, a large number of intelligently adjustable reflecting elements at the IRS can focus the signal energy into a sharp beam toward the user, which achieves high beamforming gain similarly as in M-MIMO, but only via a passive array with significant energy saving.  In general, the transmit beamforming  at the AP needs to be jointly designed with the phase shifts at the IRS based on all the AP-IRS, IRS-user, and AP-user channels  in order to fully reap their beamfroming gains. However, the formulated optimization problem is shown to be non-convex and difficult to be solved optimally.

  To tackle the non-convexity of the considered problem, we first propose a centralized algorithm based on the technique of semidefinite relaxation (SDR) to obtain both a performance upper bound and a high-quality approximate solution. Such a centralized implementation requires the global channel state information (CSI) available at the IRS, and thus incurs excessive channel estimation and signal exchange  overheads at/between the AP and IRS. To reduce such overheads and achieve low complexity,  we further propose a distributed algorithm inspired by the alternating optimization. The key idea is that the AP and  IRS independently adjust the transmit beamforming and phase shifts in an alternating manner until the convergence is reached.
It is shown by simulation  that the link signal-to-noise ratio (SNR)   can be significantly improved by deploying the IRS as compared to the conventional setup without the IRS.
In addition, with the proposed beamforming design, it is shown that the receive SNR in the vicinity of the IRS increases with the number of its reflecting elements $N$ in the order of  $N^2$, which implies that significant power saving at the AP or SNR gain at the user can be achieved in practice.

\section{System Model and Problem Formulation}
\subsection{System Model}
As shown in Fig. \ref{system:model}, we consider a point-to-point multiple-input single-output (MISO) wireless system where an AP equipped with $M$ antennas serves a single-antenna user. To enhance the link performance, an IRS composed of $N$ passive elements  is installed on  a surrounding wall to assist in the AP-user communication/power transfer.  Equipped with a smart controller, the IRS can dynamically adjust the phase shift of each reflecting  element based on the propagation environment learned through periodic sensing via the same passive array (when not reflecting). In particular, IRS controller coordinates the switching between two working modes, i.e., receiving mode for environment sensing  (e.g., CSI estimation) and reflecting mode for scattering the  incident signals from the AP \cite{subrt2012intelligent}.
Due to significant path loss, it is assumed that the power of the signals that are reflected by the IRS two or more times  is negligible and thus ignored.
 In addition, we assume a quasi-static flat-fading  channel model for all channels involved in our considered setup. Although we focus on the downlink communication from the AP to the user, the results are also applicable to the uplink. Since the IRS is a passive reflecting device,  we consider  a time-division duplexing (TDD) protocol for the uplink and downlink transmissions and exploit channel reciprocity for the CSI acquisition at the IRS in both link directions.
%

The baseband equivalent channels of the  AP-user link, IRS-user link, and AP-IRS link are denoted by $\bm{h}^H_d\in \mathbb{C}^{1\times M}$,  $\bm{h}^H_r\in \mathbb{C}^{1\times N}$, and $\bm{G}\in \mathbb{C}^{N\times M}$, respectively, where  the  superscript  $H$ represents the conjugate transpose operation and
$\mathbb{C}^{a\times b}$ denotes the space of $a\times b$ complex-valued matrices.  It is worth noting that the indirect  channel from the AP to user via IRS  is also referred to as dyadic backscatter channel or pinhole/keyhole channel in the literature \cite{griffin2009complete}, which behaves quite different from the AP-user direct channel. Specifically, each element on the IRS behaves like a pinhole/keyhole, which combines all the received multi-path signals  at a single physical point, and re-scatters the combined signal as if from a point source. Let $\bm{\theta}= [\theta_1, \cdots, \theta_N]$  and $\Theta = \text{diag} (\beta e^{j\theta_1}, \cdots, \beta e^{j\theta_n}, \cdots,  \beta e^{j\theta_N})$  (with $j$ denoting the imaginary unit and  $\text{diag}(\mathbf{a})$ denoting a diagonal matrix with each diagonal element being the corresponding element in $\mathbf{a}$) denote the diagonal phase-shift matrix for the IRS, where $\theta_n\in [0, 2\pi]$ and $\beta \in [0, 1]$\footnote{In practice, each element on the IRS is designed to maximize signal reflection. Thus, we set $\beta=1$ in the sequel of this paper. Note that this scenario is different from backscatter communications where the RFID tags need to harvest a certain amount of energy from the incident signals for circuit operations and thus have a much smaller amplitude reflection coefficient in practice.} are the phase shift and amplitude reflection coefficient on the combined incident signal, respectively.
Then, the composite AP-IRS-user channel  can be modeled as a concatenation of three components, namely, AP-IRS link, IRS reflecting with phase shifts, and IRS-user link. Therefore, it differs from the conventional AF relay channel since the relay amplifies not only its received source signal but also its own receiver noise, and forwards the amplified (as opposed to ``reflected'')  signal to the destination. In this paper, we consider linear beamforming at the AP, with  $\bm{w}\in \mathbb{C}^{M\times 1}$  denoting the transmit beamforming vector. Denote by $\bar p$ the maximum transmit power at the AP, i.e., $\|\bm{w}\|^2\leq \bar p$, where  $\|\cdot\|$ denotes the Euclidean
norm of  a complex vector.
 Then, the total signal received at the user can be expressed as
\begin{align}\label{SectionII:channel}
y= ( \bm{h}^H_r\Theta \bm{G} +  \bm{h}^H_d)\bm{w}s + z,
\end{align}
where $s$ is independent and identically distributed (i.i.d.) random variable with zero mean and unit variance, and $z$ denotes additive white Gaussian noise (AWGN) at the user receiver with zero mean and variance $\sigma^2$. Accordingly, the signal power received at the user is given by
\begin{align}\label{SectionII:receivedpower}
\gamma = {|( \bm{h}^H_r\Theta \bm{G}+\bm{h}^H_d)\bm{w} |^2}.
\end{align}

\subsection{Problem Formulation}
In practice, the proposed system can be applied to either wireless power or information transfer. In the former case, the harvested energy is generally modeled as a concave and increasing  function of the received signal power $\gamma$. In the latter case, the information achievable rate is a logarithmic function of the receive SNR, which also increases with $\gamma$. Thus, in this paper, we focus on maximizing the received signal power in \eqref{SectionII:receivedpower} by jointly optimizing the transmit beamforming $\bm{w}$ and the phase shifts $\bm{\theta}$, subject to the maximum transmit power constraint at the AP.
The corresponding optimization problem can be formulated as
\begin{align}
\text{(P1)}: ~~\max_{\bm{w}, \bm{\theta}} ~~~&| (\bm{h}^H_r\Theta \bm{G}+\bm{h}^H_d )\bm{w}|^2  \label{eq:obj}\\
\mathrm{s.t.}~~~~& \|\bm{w}\|^2\leq \bar p,  \\
& 0\leq \theta_n \leq 2\pi, \forall\, n=1,\cdots, N. \label{phase:constraints}
\end{align}
Although all the constraints are convex, problem (P1) is a non-convex optimization problem due to the non-concave objective function with respect to  $\bm{w}$ and $\bm{\theta}$. In general, there is no standard method for solving such non-convex optimization problems optimally. In the next two sections, we propose a centralized algorithm as well as a distributed algorithm for solving (P1) by applying the SDR and alternating optimization techniques, respectively.

\section{Centralized Algorithm}
In this section, we first apply SDR to solve problem (P1) by assuming that the global CSI is available at the IRS. Then, a centralized algorithm   is proposed for implementing this solution.

For any given phase shifts $\bm{\theta}$, it can be verified that the maximum-ratio transmission (MRT) is the optimal transmit beamforming solution  to problem (P1) \cite{tse2005fundamentals}, i.e.,
$\bm w^* = \sqrt{ \bar p} \frac{(\bm{h}^H_r\Theta \bm{G}+\bm{h}^H_d  )^H}{\|\bm{h}^H_r\Theta \bm{G} +\bm{h}^H_d \|}\triangleq \bm{w}_{\rm MRT}$. By substituting $\bm w^*$ to \eqref{eq:obj},  (P1)  can be simplified as the following equivalent problem
\begin{align}\label{secIII:p2}
\text{(P2)}: ~~\max_{\bm{\theta}} ~~~&\|\bm{h}^H_r\Theta \bm{G}+ \bm{h}^H_d\|^2\\
\mathrm{s.t.}~~~~& 0\leq \theta_n \leq 2\pi, \forall\, n=1,\cdots, N.  \label{SecIII:phaseconstraint}
\end{align}
Let $\bm{v} = [v_1, \cdots, v_N]^H$ where $v_n = e^{j\theta_n}$, $\forall\, n$. Then, constraints in \eqref{SecIII:phaseconstraint} are equivalent to $|v_n|=1, \forall\, n=1,\cdots, N$. By applying the change of variables $\bm{h}^H_r\Theta \bm{G} =\bm{v}^H\bm{\Phi} $ where $\bm{\Phi}=\text{diag}(\bm{h}^H_r)\bm{G}$, we have $\|\bm{h}^H_r\Theta \bm{G} + \bm{h}^H_d\|^2 =\|\bm{v}^H\bm{\Phi}+ \bm{h}^H_d\|^2 $. Thus, problem (P2) is equivalent to
\begin{align}
\text{(P3)}: ~~\max_{\bm{v}} ~~~&\bm{v}^H\bm{\Phi}\bm{\Phi}^H\bm{v} + \bm{v}^H\bm{\Phi}\bm{h}_d+\bm{h}^H_d\bm{\Phi}^H \bm{v} \\
\mathrm{s.t.}~~~~& |v_n|=1, \forall\, n=1,\cdots, N. \label{P4:C9}
\end{align}
Problem (P3) is a non-convex quadratically constrained quadratic program (QCQP), which can be reformulated as a homogeneous QCQP. Specifically,
by introducing an auxiliary variable $t$, problem (P3) can be equivalently written as
\begin{align}
\text{(P4)}: ~~\max_{\bm{\bar{v}}} ~~~&\bm{\bar{v}}^H\bm{R}\bm{\bar{v}}  \\
\mathrm{s.t.}~~~~& |\bar{v}_n|=1, \forall\, n=1,\cdots, N+1, \label{P4:C9}
\end{align}
where
\[
\bm{R}=\begin{bmatrix}
\bm{\Phi}\bm{\Phi}^H  & \bm{\Phi}\bm{h}_d \\
\bm{h}^H_d\bm{\Phi}^H  & 0 \\
\end{bmatrix},~~
\bm{\bar{v}}=\begin{bmatrix}
\bm{v}  \\
t \\
\end{bmatrix}.
\]
However, problem (P4) is NP-hard in general \cite{so2007approximating}. 
Note that $\bm{\bar{v}}^H\bm{R}\bm{\bar{v}}={\rm{tr}}(\bm{R}\bm{\bar{v}}\bm{\bar{v}}^H)  $.
Define $\bm{V}=\bm{\bar{v}}\bm{\bar{v}}^H$, which needs to satisfy  $\bm{V}\succeq \bm{0}$ and ${\rm{rank}}(\bm{V})=1$. Since the rank-one constraint is non-convex, we apply SDR to relax this constraint. As a result,
problem (P4) is reduced to
\begin{align}
\text{(P5)}: ~~\max_{\bm{V}} ~~~&{\rm{tr}}(\bm{RV})  \\
\mathrm{s.t.}~~~~& \bm{V}_{n,n} = 1, \forall\, n=1,\cdots, N+1, \label{P6:C9} \\
~~~~&\bm{V} \succeq 0.  \label{P6:C9}
\end{align}
It can be observed that problem (P5) is a standard convex semidefinite program (SDP) and hence it can be optimally solved by existing convex optimization solvers such as CVX \cite{cvx}.
Generally,  the relaxed problem (P5) may not lead to a rank-one solution, i.e., ${\rm{rank}}(\bm{V})\neq1$, which implies that  the optimal objective value of (P5) only serves an upper bound of (P4). Thus, additional steps are needed to construct  a rank-one solution from the optimal higher-rank solution to problem (P5). Specifically, we first obtain  the eigenvalue decomposition of $\bm{V}$ as $\bm{V}= \bm{U\Sigma U}^H$, where $\bm{U}= [e_1, \cdots, e_{N+1}]$ and $\bm{\Sigma} =
\text{diag}(\lambda_1, \cdots , \lambda_{N+1})$ are a  unitary matrix and a diagonal matrix, respectively, both with the size of $(N+1)\times(N+1)$. Then, we obtain a suboptimal solution to (P4) as  $\bm{\bar v} = \bm{U\Sigma^{1/2}r}$, where $\bm{r} \in  \mathbb{C}^{(N+1)\times 1}$ is a random vector generated according to  $\bm{r} \in \mathcal{CN}(\bm{0}, \bm{I}_{N+1})$ with $\mathcal{CN}(\bm{0}, \bm{I}_{N+1})$ denoting the the circularly symmetric complex
Gaussian (CSCG) distribution with zero mean  and covariance matrix $\bm{I}_{N+1}$.  With independently generated Gaussian random vectors $\bm{r}$, the objective value of (P4) is approximated as the maximum one attained by the best  $\bm{\bar v}$ among all $\bm{r}$'s.
Finally,  the solution $\bm{v}$ to problem (P3) can be recovered by  $ \bm{v} =e^{j\arg([ \frac{ \bm{\bar v}}{{\bar v}_{N+1}}]_{(1:N)})}$ where $[\bm{x}]_{(1:N)}$ denotes the vector that contains the first $N$ elements in $\bm{x}$.  It has been shown that such an SDR approach followed by sufficiently large number of randomizations of $\bm{r}$   guarantees an $\frac{\pi}{4}$-approximation of the optimal objective value of problem (P3) \cite{so2007approximating}.

To implement the above solution, the following centralized algorithm is proposed. First,  the user sends a pilot signal while both the AP and IRS perform channel estimation to obtain $\bm{h}^H_d$ and $\bm{h}^H_r$, respectively. Second, the AP sends a pilot signal while the IRS performs channel estimation to obtain $\bm{G}$. Third, the AP sends the obtained CSI knowledge of $\bm{h}^H_d$ to the IRS via a dedicated wireless control link, as shown in Fig. \ref{system:model}. With the above protocol, the CSI of all links is available at the IRS and thus it can compute $\bm{w}$ and $\bm{\theta}$ by solving problem (P1).\footnote{We assume that  the centralized algorithm is performed at the IRS rather than the AP since otherwise the estimated channel matrix  $\bm{G}$ needs to be fed back to the AP from the IRS, which incurs more feedback overheads.}
Finally, the IRS controller sends the optimized transmit beamforming $\bm{w}$ back to the AP for transmission.  Although the above centralized design achieves the best performance, its required channel estimation and signal exchange  overheads may be prohibitive in practice, especially when the number of antennas at the AP and/or that of elements at the IRS are large.

\section{Distributed Algorithm}

To facilitate practical implementation, we propose a low-complexity distributed  algorithm in this subsection based on alternating optimization.  Specifically, the transmit beamforming at the AP and the phase shifts at the IRS are optimized iteratively in an alternating manner with one being fixed in each iteration, until both reach the convergence or a maximum number of iterations is executed in practice. It is worth pointing out that the alternating optimization itself does not necessarily imply  a distributed implementation, while exploiting the special structure of our formulated problem enables us to avoid channel feedback/signal exchange between the AP and IRS and also reduce the channel estimation complexity as compared to the centralized algorithm. We first present the alternating optimization based method to solve (P1) as follows.

For any given transmit beamforming $\bm{w}$,  the objective function of (P1) satisfies the following inequality:
\begin{align}\label{SecIV:distributed}
| (\bm{h}^H_r\Theta \bm{G}+\bm{h}^H_d )\bm{w}| &=| \bm{h}^H_r\Theta \bm{G}\bm{w}+ \bm{h}^H_d\bm{w}| \nonumber\\
& \overset{(a)} \leq  | \bm{h}^H_r\Theta \bm{G}\bm{w}| + | \bm{h}^H_d\bm{w}|,
\end{align}
where the equality in $(a)$ holds if and only if  $\mathrm{arg}(\bm{h}^H_r\Theta \bm{G}\bm{w})=\mathrm{arg}(   \bm{h}^H_d\bm{w} )\triangleq \varphi_0$, with  $\mathrm{arg}(\cdot)$ denoting  the component-wise phase of a complex vector. Next, we show that there always exists a solution $\bm{\theta}$ that satisfies $(a)$ with equality as well as  the phase shifter constraints in \eqref{phase:constraints}.
By applying the change of variables  $\bm{h}^H_r\Theta\bm{G}\bm{w} =\bm{v}^H\bm{a}$ where $\bm{v} = [e^{j\theta_1}, \cdots, e^{j\theta_N}]^H$ and $\bm{a}=\text{diag}(\bm{h}^H_r)\bm{G}\bm{w}$, problem (P1) is reduced to (by ignoring the constant term $| \bm{h}^H_d\bm{w}|$)
\begin{align}
\text{(P1')}: ~~\max_{\bm{v}} ~~~&|\bm{v}^H\bm{a}|\\
\mathrm{s.t.}~~~~&  |v_n|=1, \forall\, n=1,\cdots, N,\\
& \mathrm{arg}(\bm{v}^H\bm{a})=  \varphi_0. 
\end{align}
It can be verified that the optimal solution to problem (P1') is given by $\bm{v}^* = e^{j (\varphi_0 -  \arg(\aaa) )}=e^{j ( \varphi_0 - \arg( \text{diag}(\bm{h}^H_r)\bm{G}\bm{w}) )}$. Thus, the corresponding $n$th phase shift at the IRS is given by
\begin{align}\label{phase:sub}
\theta^*_n &=\varphi_0 -  \arg({h}^H_{n,r}\bm{g}^H_{n}\bm{w}) \nonumber\\
&= \varphi_0 -  \arg({h}^H_{n,r})- \arg(\bm{g}^H_{n}\bm{w}),
\end{align}
where ${h}^H_{n,r}$ is the $n$th element of $\bm{h}^H_r$ and $\bm{g}^H_{n}$ is the $n$th row vector of $\bm{G}$. Note that $\bm{g}^H_{n}\bm{w}$ combines the transmit beamforming and the AP-IRS channel, and thus can be regarded as the equivalent channel perceived by the $n$th reflecting element at the IRS.
 Therefore, \eqref{phase:sub} suggests that the $n$th phase shift should be set such that the phase of the signal that passes through the AP-IRS and IRS-user links is aligned with that of the signal over the  AP-user direct  link to achieve coherent signal construction at the user. Furthermore, it is interesting to note that the obtained phase $\theta^*_n $ is independent of the amplitude of ${h}_{n,r}$.

Next, we optimize the transmit beamforming for given $\bm{\theta}$.  As in Section III, the composite AP-user channel is given by $\bm{h}^H_r\Theta \bm{G} +  \bm{h}^H_d$ and hence MRT is optimal, i.e., $\bm{w}_{\rm MRT} =\sqrt{ \bar p} \frac{(\bm{h}^H_r\Theta \bm{G}+\bm{h}^H_d  )^H}{\|\bm{h}^H_r\Theta \bm{G} +\bm{h}^H_d \|}$.   Recall that  $\mathrm{arg}(   \bm{h}^H_d\bm{w} )$(i.e., $\varphi_0$) is needed for solving problem (P1'). However,  $\mathrm{arg}(\bm{h}^H_d\bm{w}_{\rm MRT})$ is generally different at different iterations, thus it needs to be
  fed back from the AP to the IRS for solving problem (P1'). To enable a distributed implementation, we constrain the phase of $\bm{h}^H_d\bm{w}$ to be a constant (e.g., 0) for all iterations.
The key observation is that  an arbitrary (common) phase rotation can be added to the beamforming vector without changing the beamforming gain. As such, the transmit  beamforming vector is modified as
\begin{align}\label{rotate:beamformer}
\bm{w}^* =\sqrt{ \bar p} \frac{(\bm{h}^H_r\Theta \bm{G}+\bm{h}^H_d  )^H}{\|\bm{h}^H_r\Theta \bm{G} +\bm{h}^H_d \|}e^{j\alpha},
 \end{align}
 where $\alpha$ is adaptively chosen by the AP in each iteration such that $\bm{h}^H_d\bm{w}^*$ is a real number, i.e., $\varphi_0 = \mathrm{arg}(\bm{h}^H_d\bm{w}^*)=0$.  As a result,  (P1') becomes a local optimization problem at the IRS where the optimal solution can be  obtained by using \eqref{phase:sub}.

The distributed algorithm to realize the above alternating optimization is given in Algorithm  \ref{Algo:whole} in detail.
 Compared to the centralized algorithm in Section III, the distributed algorithm has the following advantages. First, there is no channel feedback needed between the AP and IRS. Second, only phases $\arg(\bm{h}^H_{r})$ and $\arg(\bm{g}^H_{n}\bm{w})$ need to be estimated at the IRS instead of the full CSI $\bm{h}^H_r$ and  $\bm{G}$.   Third, there is no need to run the SDP solver since closed-form solutions are available.

 \begin{algorithm}[t]
\caption{Distributed Algorithm.}\label{Algo:whole}
\begin{algorithmic}[1]
\STATE Initialize $\epsilon$ and iteration number $k=1$.
\STATE Let the IRS be in the receiving mode and the user broadcast a pilot signal. The IRS estimates phase $\arg(\bm{h}^H_{r})$ and the AP estimates the channel $\bm{h}_{d}$. The AP sets transmit beamforming as $\bm{w}^k = \sqrt{\bar p}\frac{\bm{h}_d}{\|\bm{h}_d\|}$.
\REPEAT
\STATE Let the IRS be in the receiving mode and the AP broadcast a pilot signal with transmit beamforming $\bm{w}^{k}$. The IRS estimates $\arg(\bm{g}^H_{n}\bm{w})$ and computes  the phase shift $\bm{\theta}^{k+1}$ by using \eqref{phase:sub} with given $\bm{w}^{k}$.
\STATE Let the IRS be in the reflecting mode with given $\bm{\theta}^{k+1}$ and the user broadcast a pilot signal.  The AP estimates the composite channel $\bm{h}^H_r\Theta \bm{G} +  \bm{h}^H_d$ and computes the new transmit beamforming $\bm{w}^{k+1}$ by using  \eqref{rotate:beamformer} with given $\bm{\theta}^{k+1}$.
\STATE Update $k=k+1$.
\UNTIL{ the fractional increase of the objective value  is below a threshold $\epsilon>0$ or  the maximum number of iterations is reached.}
\end{algorithmic}
\end{algorithm}

%

\section{Simulation Results}
 We consider a uniform linear array (ULA) at the AP and a uniform rectangular array (URA) at the IRS with $N=N_{x}N_{y}$ where $N_{x}$ and $N_{y}$ denote the number of reflecting elements in the horizontal and vertical directions, respectively. Specifically, the ULA is positioned  in parallel to the URA at the same altitude. For the purpose of exposition, we fix $N_y=10$ and increase $N_x$ linearly with $N$.
 The signal attenuation at a reference distance of 1 meter (m) is set as 30 dB  for all channels. Furthermore, it is assumed that the AP-IRS channel is dominated by the LoS link since in practice the IRS is usually deployed with the knowledge of the AP's location to exploit LoS channel. As a result, the  corresponding channel matrix  $\bm{G}$  is of rank one in which the row/column vectors are linearly dependent. On the other hand,  due to the user's mobility and the complex propagation environment (such as indoor),  we consider 10 dB penetration loss for both the (direct) AP-user and IRS-user channels with independent  Rayleigh fading and the pathloss exponent of 3.  It is assumed that the antenna gain of both the AP and user is 0 dBi and that of each reflecting element at the IRS is 5 dBi \cite{griffin2009complete}.
For all the simulations, we consider an information transmission scenario and use the SNR at the user receiver as the performance metric. Other required parameters are set as follows: $\epsilon=10^{-4}$,  $\sigma^2=-80$\,dBm, $\bar p=5$\,dBm, $M=8$, and $N_{x}=5$ (if not specified otherwise).

\begin{figure}[!t]
\centering
\includegraphics[width=0.35\textwidth]{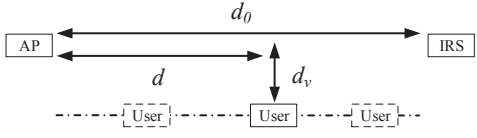}
\caption{Simulation setup. } \label{simulation:setup}\vspace{-5mm}
\end{figure}

\subsection{SNR versus AP-User Distance}
\begin{figure}[!t]
\centering
\includegraphics[width=0.39\textwidth]{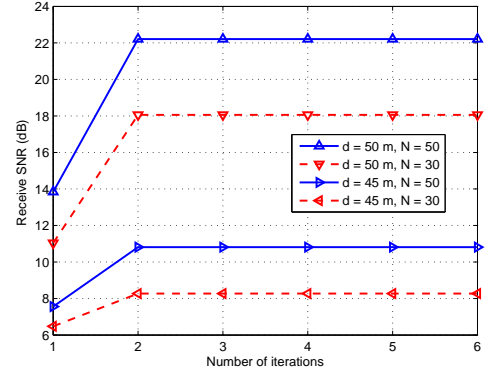}
\caption{Convergence of the proposed distributed algorithm.} \label{convergence} \vspace{-3mm}
\end{figure}

As shown in Fig. \ref{simulation:setup}, we consider the setup where the AP and IRS are located $d_0= 51$ m apart and the user lies on a horizontal line that is in parallel to the one that connects the AP and IRS, with the vertical distance between these two lines being  $d_v = 2$ m.
Denote the horizontal distance between the AP and  user  by $d$. Accordingly, the AP-user and IRS-user link distances are given by $d_1= \sqrt{d^2+d_v^2}$ m and $d_2= \sqrt{(d_0-d)^2+d_v^2}$ m, respectively. As such, we can study the receive SNR of a user located between the AP and  IRS by varying the value of $d$ m.
 In Fig. \ref{convergence}, we first show the convergence of the proposed distributed algorithm. It is observed that for all setups, at most three iterations are needed for convergence, which suggests a low implementation complexity. For performance comparison, we consider the following  schemes for simulations: 1) Upper bound: the optimal objective value of (P5); 2) Joint AP and IRS beamforming design by the  centralized algorithm proposed  in Section III; 3) Joint AP and IRS beamforming design by the  distributed algorithm proposed   in Section IV; 4) AP-user MRT: we set $\bm{w} = \sqrt{\bar p}\frac{\bm{h}_d}{\|\bm{h}_d\|}$ to achieve MRT based on the AP-user direct channel;
 5) AP-IRS MRT: we set $\bm{w} = \sqrt{\bar p}\frac{\bm{g}_n}{\|\bm{g}_n\|}$ to achieve MRT based on the AP-IRS rank-one channel; 6) Benchmark scheme without the IRS by setting  $\bm{w} = \sqrt{\bar p}\frac{\bm{h}_d}{\|\bm{h}_d\|}$. Note that for schemes 4) and 5) with given $\bm{w}$, the phase shifts are optimized by using \eqref{phase:sub}.

\begin{figure}[!t]
\centering
\includegraphics[width=0.39\textwidth]{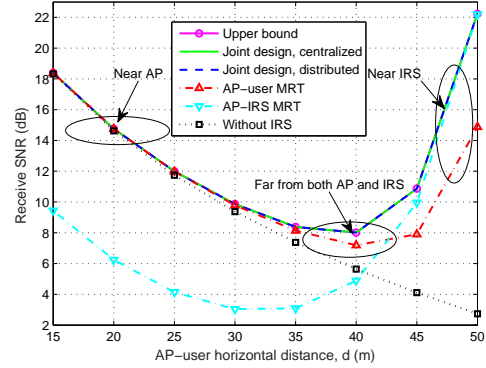}
\caption{Receive SNR versus AP-user horizontal distance, $d$. } \label{simulation:distance} \vspace{-4mm}
\end{figure}

In Fig. \ref{simulation:distance}, we compare the receive SNR of all schemes versus the horizontal distance between the AP and user, $d$. First, it is observed that the proposed centralized and distributed algorithms both achieve near-optimal SNR as compared to the performance upper bound and also significantly outperform other benchmark schemes. Second, for the scheme without the IRS, one can observe that  the user farther away from the AP suffers more SNR loss due to signal attenuation.
  However, this problem is alleviated by deploying an IRS, which implies that a larger AP-user distance does not necessarily lead to a worse user SNR in IRS-enhanced wireless networks.
This is because the user farther away from the AP is closer to the IRS and thus  it is able to receive stronger reflected signals from the IRS. As a result, the user close to either the AP (e.g., $d=17$ m)  or IRS (e.g., $d=47$ m) can achieve better SNR than a user far away from both of them (e.g., $d=40$ m).
  This result suggests that the signal coverage can be effectively extended by deploying only a passive IRS rather than installing an additional AP or active relay. For example, for a target SNR of 8 dB, the coverage of the network without the IRS is about 33 m whereas this value is improved to about 50 m by applying the proposed joint beamforming designs with an IRS.

 On the other hand, it is also observed from Fig. \ref{simulation:distance} that the AP-user MRT scheme performs close to optimal when the user is nearer to the AP, while it results in  considerable SNR loss when the user is nearer to the IRS. This is expected since in the former case, the user received signal is dominated by the AP-user direct   link whereas the IRS-user link is dominant in the latter case. Moreover, it can be observed that the AP-IRS MRT scheme behaves oppositely as the user moves away from the AP towards IRS. This further demonstrates that the proposed joint beamforming designs  can dynamically adjust the AP's beamforming to strike an optimal  balance between the signal power to  the user and IRS, respectively.

\begin{figure}[!t]
\centering
\subfigure[$d = 15$ m and $d = 50$ m]{\includegraphics[width=0.39\textwidth]{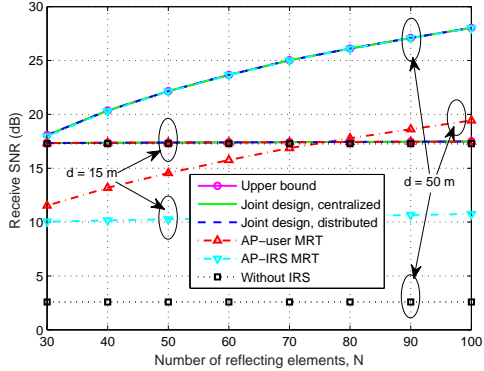}} 
\subfigure[$d = 43$ m]{\includegraphics[width=0.39\textwidth]{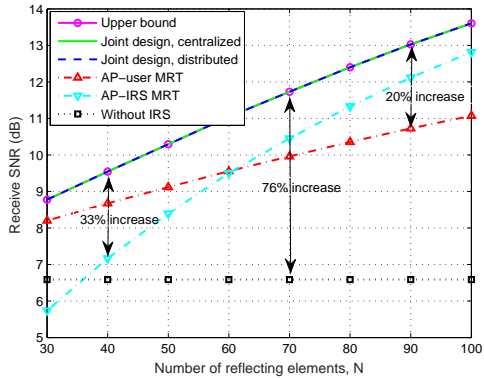}} 
\caption{Receive SNR versus number of reflecting elements at the IRS, $N$. } \label{simulation:N} \vspace{-4mm}
\end{figure}
\subsection{SNR versus Number of Reflecting Elements}

In Fig. \ref{simulation:N}, we compare the receive  SNR of all schemes versus the number of reflecting elements at the IRS when $d=15$, $43$, and $50$ m, respectively. From Fig. \ref{simulation:N} (a), it is first observed that for the case of $d=50$ m, the AP-IRS MRT scheme achieves near-optimal SNR since the signal reflected by the IRS is much stronger than that directly from the AP at the user. Furthermore, it is interesting to note that the user receive SNR of the proposed schemes scales with the number of reflecting elements $N$ in the order of  $N^2$. For example, when $N=30$, the user is able to achieve an SNR of 18 dB and this value is improved to 24 dB when  $N=60$, which suggests a 6 dB gain by doubling the number of reflecting elements. Such a performance gain is attributed to two aspects. On one hand, increasing $N$ enables more reflecting elements to receive the signal energy from the AP, which leads to an array gain of $N$. On the other hand, the passive beamforming by the phase shifters  achieves another reflect  beamforming gain of $N$. Thus, a total beamforming gain of $N^2$ is resulted. Note that in contrast, the conventional MRT beamforming with $N$ active antennas only achieves a beamforming gain of $N$.
  Therefore, the proposed IRS can take advantage of both the large-scale aperture gain and reflect beamforming gain with only passive phase shifters, which is thus more spectral and energy efficient. Finally, it is observed that this gain diminishes as the user moves away from the IRS. For example, for the case of $d=15$ m where the AP-user direct link signal is much stronger than that of IRS-user link, the user receive SNR is not sensitive to the number of reflecting elements.
For the case of $d=43$ m as shown in Fig. \ref{simulation:N} (b) when the user is neither close to the AP nor close to the IRS, it is observed that the SNR gain is generally lower than $N^2$. This is because in this case the signal energy received at the IRS is compromised as the AP transmit beamforming is steered to strike a balance between the AP-IRS link and the AP-user direct link. In practice, the number of reflecting elements can be properly selected depending on the IRS's location as well as the target user SNR/AP coverage.

\section{Conclusion}
In this paper, we have proposed a new approach to enhance the performance  of wireless networks by deploying passive IRSs. Specifically, the active  transmit beamforming at the AP and the passive reflect beamforming of the phase shifters at the IRS  are jointly optimized to maximize  the signal power received at the user in an IRS-enhanced point-to-point MISO system.
By leveraging the SDR and alternating optimization techniques, respectively, we propose both  centralized and distributed designs. In particular, the low-complexity distributed design has been shown to achieve near-optimal performance and is thus appealing for practical implementation. Simulation results also demonstrate the SNR improvement and signal coverage extension achieved by deploying the IRS as compared to the conventional setup without the IRS. The proposed joint AP and IRS beamforming designs are also shown to be effective and crucial to achieve optimal performance under different setups.


\bibliographystyle{IEEEtran}
\bibliography{IEEEabrv,mybib}

\end{document}